\documentclass[12pt,letter]{article}

\usepackage{graphicx, epsfig, color}
\def\eslt{E_T^{\rm miss}}
\def\to{\rightarrow}

\def\bi{\begin{itemize}}
\def\ei{\end{itemize}}

\def\tst{\tilde t}
\def\ttau{\tilde \tau}

\def\tg{\tilde g}

\def\tell{\tilde\ell}
\def\tq{\tilde q}
\def\tw{\widetilde W}
\def\tz{\widetilde Z}
\def\alt{\stackrel{<}{\sim}}
\def\agt{\stackrel{>}{\sim}}
\def\be{\begin{equation}}  
\def\ee{\end{equation}}  
\def\bea{\begin{eqnarray}}  
\def\eea{\end{eqnarray}}  
\def\beas{\begin{eqnarray*}}  
\def\eeas{\end{eqnarray*}}  
\newcommand\prd[3]{{\it Phys.\ Rev.\ }{\bf D #1} (#2) #3}

\newcommand\prl[3]{{\it Phys.\ Rev.\ Lett.\ }{\bf #1} (#2) #3}
\newcommand\plb[3]{{\it Phys.\ Lett.\ }{\bf B #1} (#2) #3}
\newcommand\jhep[3]{{\it J. High Energy Phys.\ }{\bf #1} (#2) #3}
\newcommand\app[3]{{\it Astropart.\ Phys.\ }{\bf #1} (#2) #3}

\newcommand\npb[3]{{\it Nucl.\ Phys.\ }{\bf B #1} (#2) #3}
\newcommand\epjc[3]{{\it Eur.\ Phys.\ J. }{\bf C #1} (#2) #3}
\newcommand\ptp[3]{{\it Prog.\ Theor.\ Phys.\ }{\bf #1} (#2) #3}
\newcommand{\hepph}[1]{hep-ph/#1}
\newcommand{\hepth}[1]{hep-th/#1}
\newcommand{\hepex}[1]{hep-ex/#1}
\newcommand{\astroph}[1]{astro-ph/#1}

\begin{document}
\begin{titlepage}
\begin{flushright}
FSU-HEP/070330\\
MADPH-07-1480
\end{flushright}

\vspace{0.5cm}
\begin{center}
{\Large \bf 
Precision gluino mass at the LHC\\ 
in SUSY models with decoupled scalars
}\\ 
\vspace{1.2cm} \renewcommand{\thefootnote}{\fnsymbol{footnote}}
{\large Howard Baer$^{1,2}$\footnote[1]{Email: baer@hep.fsu.edu },
Vernon Barger$^2$\footnote[2]{Email: barger@physics.wisc.edu}, 
Gabe Shaughnessy$^2$\footnote[3]{Email: gshau@hep.wisc.edu}, \\
Heaya Summy$^1$\footnote[4]{Email: heaya.summy@gmail.com}, 
Lian-Tao Wang $^3$\footnote[5]{Email: lianwang@princeton.edu}} \\
\vspace{1.2cm} \renewcommand{\thefootnote}{\arabic{footnote}}
{\it 
1. Dept. of Physics,
Florida State University, Tallahassee, FL 32306, USA \\
2. Dept. of Physics,
University of Wisconsin, Madison, WI 53706, USA \\
3. Department of Physics, Princeton University, Princeton, NJ 08543\\
}

\end{center}

\vspace{0.5cm}
\begin{abstract}
\noindent 
One way to ameliorate the SUSY flavor and CP problems is to postulate that scalar masses
lie in the TeV or beyond regime.
For example, the focus point (FP) region of the minimal supergravity (mSUGRA) model is especially
compelling in that heavy scalar masses can co-exist with low fine-tuning while 
yielding the required relic abundance of cold dark matter
(via a mixed higgsino-bino neutralino). We examine many of the characteristics of
collider events expected to arise at the CERN LHC in models with multi-TeV scalars, taking the
mSUGRA FP region as a case study. The collider events are
characterized by a hard component arising from gluino pair production, 
plus a soft component arising from direct chargino and neutralino production. 
Gluino decays in the FP region are
characterized by lengthy cascades yielding very large jet and lepton multiplicities,
and a large $b$-jet multiplicity.
Thus, as one steps to higher jet, $b$-jet or lepton multiplicity, signal-over-background
rates should steadily improve.
The lengthy cascade decays make mass reconstruction via kinematic edges difficult;
however, since the hard component is nearly pure
gluino pair production, the gluino mass can be extracted to $\pm 8\%$ via total rate
for $\eslt +\ge 7$-jet $+\ge 2\ b$-jet events, assuming 100 fb$^{-1}$ of integrated
luminosity.
The distribution of invariant mass of opposite-sign/same-flavor dileptons in the 
hard component exhibits two dilepton mass edges: $m_{\tz_2}-m_{\tz_1}$ and $m_{\tz_3}-m_{\tz_1}$.
As a consistency check, the same mass edges should be seen in isolated
opposite-sign dileptons occurring in the soft component trilepton signal which originates
mainly from chargino-neutralino production.

\vspace*{0.8cm}
\noindent PACS numbers: 14.80.Ly, 12.60.Jv, 11.30.Pb, 13.85.Rm

\end{abstract}


\end{titlepage}

\section{Introduction}
\label{sec:intro}

The minimal supergravity (mSUGRA) model\cite{sugra,msugra,wss} is a 
well-motivated supersymmetric model with a small
parameter space that forms a template for many investigations of the 
phenomenological consequences of weak scale supersymmetry. In mSUGRA, it is assumed that
the minimal supersymmetric standard model, or MSSM, is a valid effective theory
of physics between the energy scales $Q=M_{GUT}$ and $Q=M_{weak}$. 
It is further assumed that all MSSM scalar masses unify to a common value $m_0$
at $M_{GUT}$, while gauginos unify to a common value $m_{1/2}$ and trilinear
soft terms unify to a common value $A_0$. The weak scale soft parameters
can be calculated by renormalization group evolution from $M_{GUT}$ to
$M_{weak}$. The large value of the top quark Yukawa coupling drives the
up-Higgs squared mass to negative values, leading to radiative electroweak 
symmetry breaking (EWSB). The EWSB minimization conditions allow one to trade 
the bilinear soft breaking term $B$ for $\tan\beta$, the ratio of Higgs field vevs.
It also determines the magnitude (but not the sign) of the superpotential
Higgs mass term $\mu$.
Thus, the entire weak scale sparticle mass spectrum and mixings can be calculated
from the well-known parameter set
\be
m_0,\ m_{1/2},\ A_0,\ \tan\beta ,\ sign(\mu ) .
\ee
Thus, once this parameter set is stipulated, a whole host of observables, 
including the neutralino dark matter relic density $\Omega_{\tz_1}h^2$ and
collider scattering events, may be calculated.   
For implementation, we use Isajet v7.74\cite{isajet,kraml} 
to calculate the sparticle mass spectrum and associated collider events, 
and IsaReD\cite{bbb} to calculate the neutralino relic density. 

One of the important consequences of the MSSM, due to $R$-parity 
conservation, is that the lightest SUSY particle (LSP) is absolutely stable.
In mSUGRA, the LSP is usually found to be the lightest neutralino $\tz_1$,
which is a weakly interacting massive particle (WIMP), and hence has the potential
to naturally match the measured abundance of cold dark matter in the universe.
An analysis of the three-year WMAP and galaxy survey data sets \cite{wmap} 
implies that the ratio of
dark matter density to critical density, 
\be
\Omega_{\tz_1}h^2\equiv \rho_{\tz_1}/\rho_c =0.111^{+0.011}_{-0.015}\ \ (2\sigma ) .
\ee
where $h=0.74\pm 0.03$ is the Hubble constant.  By comparing the mSUGRA predicted value of $\Omega_{\tz_1}h^2$ to this measured
value, one finds that only certain parts of the mSUGRA parameter space are
cosmologically allowed. These include the following.
\bi
\item The bulk region at low $m_0$ and low $m_{1/2}$, where
neutralino annihilation is enhanced by light $t$-channel 
slepton exchange\cite{bulk}.
The tight WMAP $\Omega_{CDM}h^2$ limit has pushed this allowed region to 
very small $m_0$ and $m_{1/2}$ values, while LEP2 limits on $m_{\tw_1}$ and 
$m_{\tell}$ (and possibly on $m_h$) have excluded these same  low values
so that almost no bulk region has survived\cite{bb3}.
\item The stau co-annihilation region occurs at very low $m_0$ but any $m_{1/2}$
values, so that $m_{\ttau_1}\simeq m_{\tz_1}$, and neutralinos can annihilate
against tau sleptons\cite{stau} in the early universe. For certain $A_0$ values
which dial $m_{\tst_1}$ to very low values, there also exists a top-squark
co-annihilation region\cite{stop}.
\item The $A$-funnel region occurs at large values of the parameter $\tan\beta\sim 50$,
where $2m_{\tz_1}\sim m_A$, and neutralinos can annihilate through the broad
pseudoscalar Higgs resonance\cite{Afunnel}. 
There is also a light Higgs resonance region where $2m_{\tz_1}\sim m_h$ at low $m_{1/2}$ values\cite{bulk,drees_h}.
\item At large $m_0$ near the boundary of parameter space, the superpotential
Higgsino mass term $\mu$ becomes quite small, and the $\tz_1$ can become
a mixed higgsino-bino neutralino. This region is known as the 
hyperbolic branch/focus point region (HB/FP)\cite{ccn,fmm,hb_fp}. 
In this case, neutralino annihilation to
vector bosons is enhanced, and a match to the WMAP measured relic density can be found. 
\ei

The HB/FP region of mSUGRA is especially compelling. In this region, the large
value of $m_0\sim$ several TeV means that possible SUSY contributions to 
various flavor-changing and $CP$-violating processes are suppressed by the large squark and slepton masses. 
For instance, SUSY contributions to the flavor-violating decay
$b\to s\gamma$ are small, so in the HB/FP region this decay rate is predicted to be 
in accord with SM predictions, as observed.
Meanwhile, the calculated amount of fine-tuning in the electroweak sector
has been shown to be small\cite{ccn,fmm}, in spite of the presence of multi-TeV
top squarks. 

In this paper, we examine the HB/FP region with regard to 
what sort of collider events are expected at the CERN LHC $pp$ collider,
which is set to begin operating in the near future at a center-of-mass
energy $\sqrt{s}=14$ TeV. Much previous work on this issue has been done.
In Ref. \cite{lhcreach,bbbkt}, the reach of the LHC in the mSUGRA model, 
including the HB/FP region, was calculated. The reach for 100 fb$^{-1}$ was found to
extend to $m_{1/2}\sim 700$ GeV, corresponding to a reach in $m_{\tg}$ of 
about 1.8 TeV. The reaches of $\sqrt{s}=0.5$ and 1 TeV $e^+e^-$ linear
colliders were also calculated\cite{bbkt}, and found to extend past that of the LHC, 
since when $\mu$ becomes small, charginos become light, and chargino pair production
is a reaction that $e^+e^-$ colliders are sensitive to, essentially up to the
kinematic limit for chargino pair production. In fact, the reach of the 
Fermilab Tevatron  for SUSY in the clean trilepton channel\cite{Barger:1998hp} is somewhat enhanced in the 
HB/FP region\cite{bkt}, since charginos and neutralinos can be quite light, 
and decay with characteristic dilepton mass edges. The reaches of
direct\cite{direct} and indirect\cite{indirect} dark matter search experiments
are also enhanced in the HB/FP region.

In Ref. \cite{hp}, Hinchliffe and Paige examined characteristic measurements
that the LHC could make for an mSUGRA sample point nearby to the HB/FP region.
They found a good signal/background ratio could be obtained with a hard cut
on effective mass $M_{eff}=\sum_{jets} E_T +\sum_{leptons} E_T+\eslt$ (e.g. $M_{eff} > 400$ GeV) and by requiring 
the presence of a $b$-jet\footnote{The effective mass was introduced and used in heavy top quark production \cite{meff-hq}.}. 
Some characteristic distributions such as $m(\ell^+\ell^-)$ which
gave a dilepton mass edge at $m_{\tz_2}-m_{\tz_1}$ and 
$m(b-jet ,\ell)<\sqrt{(m_t^2-M_W^2)/2}$
(indicating the presence of a $t$-quark in the decay chain) could be made.

In Ref. \cite{mmt}, Mizukoshi, Mercadante and Tata found that the LHC
reach in the HB/FP region could be enhanced by up to 20\% by requiring 
events with the presence of one or two tagged $b$-jets.
In Ref. \cite{bkpu}, a model-independent exploration of the HB/FP region was 
made with regard to collider and dark matter signals. 
The LHC reach via multilepton cascade decays was compared to the LHC reach via
clean trileptons from $pp\to \tw_1\tz_2\to 3\ell+\eslt$ production.
In the latter process, backgrounds from $W^*Z^*$ and $W^*\gamma^*$ were
calculated, and the trilepton reach was found to be comparable to-- but slightly
smaller than-- the reach via a search for gluino cascade decays.
In Ref. \cite{baltz}, the authors examined what sort of cosmological
measurements could be made in several mSUGRA case studies (including the point\footnote{The parameters of the point are $m_0=3280$ GeV, $m_{1/2} = 300$ GeV, $\tan \beta = 10$, $A_0=0$ GeV and $\mu > 0$.} LCC2
in the HB/FP region) by measurements at the LHC and a $\sqrt{s}=0.5$ and
1 TeV ILC. For LCC2 at the LHC, they assumed the $\tz_3-\tz_1$ and
$\tz_2-\tz_1$ mass edges could be measured to an accuracy of 1 GeV, while
it was conjectured that $m_{\tg}$ and $m_{\tz_1}$ could be measured to $\sim 10\%$ accuracy
via some kinematic distributions.

In this paper, in anticipation of the LHC turn on, 
we wish to understand many of the characteristics of collider 
events expected in the HB/FP region, with an eye towards
sparticle mass measurements rather than reach studies. We find that
the expected collider events in the HB/FP region separate themselves into
a hard component, arising from gluino pair production, and a soft component,
arising from pair production of charginos and neutralinos. The gluino pair
production events typically involve lengthy cascade decays to top
and bottom quarks \cite{sscgaugino}, and so high jet, $b$-jet and isolated lepton multiplicities
are expected. However, the complex cascade decays do not lend themselves to
simple kinematic measurements of the gluino or neutralino masses, mainly due
to the combinatorics of picking out the correct gluino decay products\footnote{Studies of gluino mass \cite{glmass} and spin \cite{glspin} determination have been made through the cascade decays $\tilde g\to b \tilde b^*_1$ with $\tilde b_1\to \tilde Z_2\to \tilde l\to \tilde Z_1$ at parameter point SPS1a.}.
We do find that the gluino mass should be extractable based on total rate in
the multi-jet+multi-lepton$+\eslt$ events to a precision of about 5-10\% for
100 fb$^{-1}$ of integrated luminosity. For both the hard and 
soft components, the $\tz_3-\tz_1$ and $\tz_2-\tz_1$ mass edges should be visible.

The remainder of this paper is organized as follows. In Sec. \ref{sec:fp},
we present details of sparticle masses and cross sections expected from the 
HB/FP region at the  LHC. In Sec. \ref{sec:evgen}, we present some details of our signal and 
background calculations.
In Sec. \ref{sec:dist} we present distributions for a variety of 
collider observables for a case study and SM backgrounds.
We present a set of cuts that allows good separation of signal vs.
background over a large range of $m_{\tg}$ values. In Sec. \ref{sec:results},
we show expected signal-to-background plots for gluino pair production and 
discuss how these can be used to extract a measurement of the gluino mass. 
In Sec. \ref{sec:leps}, we address leptonic signals.
We conclude in Sec. \ref{sec:conclude}.

\section{Sparticle production and decay in the HB/FP region}
\label{sec:fp}

In the mSUGRA model, for a given set of GUT scale soft SUSY breaking 
(SSB) masses, the associated weak scale values may be computed via 
renormalization group (RG) evolution \cite{Barger:1992ac}.
Once the weak scale SSB terms have been obtained, then the scalar
potential must be minimized to determine if electroweak symmetry
is properly broken.
While  one EWSB condition allows the bilinear parameter $B$ to be traded for $\tan\beta$,
the other condition reads (at one-loop) 
\be
\mu^2 =\frac{m_{H_d}^2-m_{H_u}^2\tan^2\beta}{(\tan^2\beta -1)}
-\frac{M_Z^2}{2} ,
\label{eq:ewsb}
\ee
which determines the magnitude of the superpotential $\mu$ parameter.
Thus, one condition that EWSB is successfully broken is that a positive value
of $\mu^2$ has been generated.
Roughly, if all the soft parameters entering Eq. \ref{eq:ewsb} are
of order $M_Z^2$, then naturalness is satisfied, and the model is not fine-tuned.

For a fixed value of the parameter $m_{1/2}$ in the mSUGRA model, 
if $m_0$ is taken to be of order the weak scale, then $m_{H_u}^2$ is driven to
negative values at the weak scale owing to the push from the large top quark Yukawa coupling 
in the RGEs.
However, if $m_0$ is taken too large, then the GUT scale value of $m_{H_u}^2$ is so high that
it is not  driven to negative
values when the weak scale is reached in RG running, 
and a positive value of $\mu^2$ cannot be found. 
Intermediate to these two extreme cases must exist a region
where $\mu^2$ is found to be zero, which forms the large $m_0$ edge
of parameter space.
If $\mu^2$ is positive, but tiny, then extremely light higgsino-like
charginos will be generated, in conflict with bounds from LEP2, 
which require $m_{\tw_1}>103.5$ GeV. If $\mu^2$ is large enough to evade LEP2 limits,
then large higgsino-bino mixing occurs in the chargino and neutralino sectors,
and in fact the lightest neutralino becomes a mixed higgsino-bino dark matter
particle. A lightest neutralino of mixed higgsino-bino form has a large annihilation 
rate to vector bosons in the early universe,
and hence may have a dark matter relic density in accord with WMAP measurements.
In this region, dubbed the hyperbolic branch/focus point region, 
multi-TeV squark and slepton masses can co-exist with low fine-tuning as dictated
by Eq. \ref{eq:ewsb}.
Thus, the HB/FP region is characterized by TeV-scale squark and slepton masses, which are
useful for suppressing possible FCNC or CP-violating processes, low fine-tuning, 
and a dark matter relic density in accord with WMAP.
Given these qualities, it is important to investigate what HB/FP supersymmetry events would look 
like at the LHC collider and what sort of mass measurements could be made in this region.

\begin{table}[htdp]
\begin{center}
\begin{tabular}{|c|ccccc|}
\hline
Point & $m_0$ & $m_{1/2}$ & $M_{\tilde g}$ & $\delta M_{\tilde g}/M_{\tilde g}$ & $\Gamma_{\tilde g}$ \\
\hline
FP0	& 2300 & 200 & 591 & LEP2 excl. & 0.2 \\
FP1	& 2450 & 225 & 655 & LEP2 excl. & 0.4 \\
FP2	& 2550 & 250 & 717 & $\pm 10\%$ & 0.6\\
FP3	& 2700 & 300 & 838 & $\pm 8\%$ & 1.1\\
FP4	& 2910 & 350 & 959 & $\pm 7\%$ & 1.8\\
FP5	& 3050 & 400 & 1076 & $\pm 8\%$ & 2.7\\
FP6	& 3410 & 500 & 1310 & $\pm 8\%$ & 5.1\\
FP7	& 3755 & 600 & 1540 & --- & 8.1\\
FP8	& 4100 & 700 & 1766 & --- & 11.8\\
FP9	& 4716 & 900 & 2211 & --- & 20.7\\
\hline
\end{tabular}
\end{center}
\caption{Points in the HB/FP region that yield a relic density $\Omega_{\tz_1}h^2\sim 0.11$.  Common values of $\tan \beta = 30$, $A_0 = 0$, $\mu >0$ and $m_t=175$ GeV are assumed.  The gluino mass in GeV and its anticipated measurement uncertainty at the LHC are also given.  The total gluino width is given in MeV.}
\label{tab:fppoints}
\end{table}%

\begin{figure}[htbp]
\begin{center}
\includegraphics[angle=-90,width=0.59\textwidth]{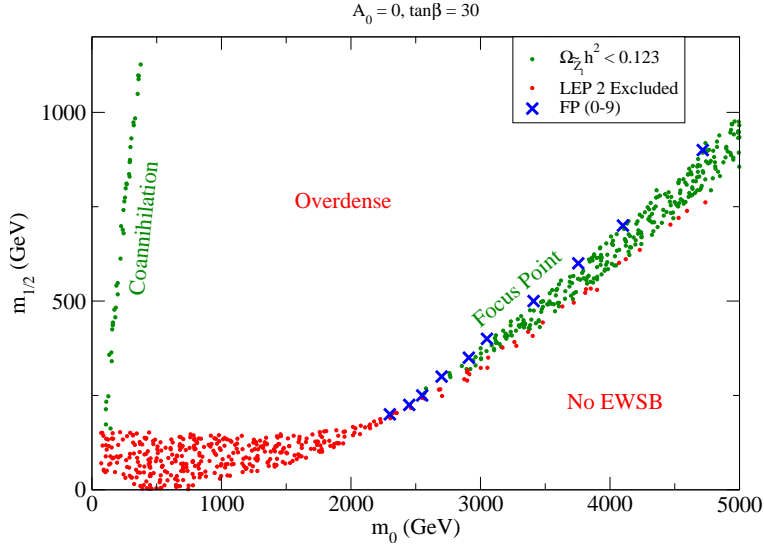}
\caption{Regions of allowed relic density in the $m_0$ vs. $m_{1/2}$ plane for $A_0 = 0$, $\tan \beta = 30$ and $m_t=175$ GeV.  Points in the HB/FP region from Table \ref{tab:fppoints} that are consistent with the observed relic density $\Omega_{\tz_1}h^2\sim 0.11$ are shown by a blue x.  The white region above the focus point and surrounding the coannihilation region give and overdensity of neutralino dark matter.}
\label{fig:rdscan}
\end{center}
\end{figure}

We have generated sparticle mass spectra in the HB/FP region using Isajet v7.74, retaining
only points which yield a relic density $\Omega_{\tz_1}h^2\sim 0.11$. These points and the anticipated measurement uncertainty are listed in Table \ref{tab:fppoints}.  Since the scalar quarks are decoupled, the gluino decays via suppressed 3-body decays, making the gluino width small and not a significant source of mass measurement uncertainty.  In Fig. \ref{fig:rdscan}, we show the points in the HB/FP from Table \ref{tab:fppoints} in the $m_0$ vs. $m_{1/2}$ plane.  Isajet uses two-loop RGEs
for the scalar mass evolution, and minimizes the RG-improved one-loop  effective potential
at an optimized scale $Q=\sqrt{m_{\tst_L}m_{\tst_R}}$ (which accounts for leading two-loop 
effects). A unique feature of Isajet's sparticle mass algorithm is that it decouples
various SSB terms from the RG evolution at their own mass scales, which gives a more
gradual transition from the MSSM to the SM effective theory, as opposed to other approaches 
which use an ``all-at-once'' transition. Thus, the Isajet algorithm should give a 
good representation of sparticle mass spectra in cases that involve a severely split mass spectrum,
such as in the HB/FP region.

\begin{figure}[htbp]
\begin{center}
\includegraphics[angle=0,width=0.59\textwidth]{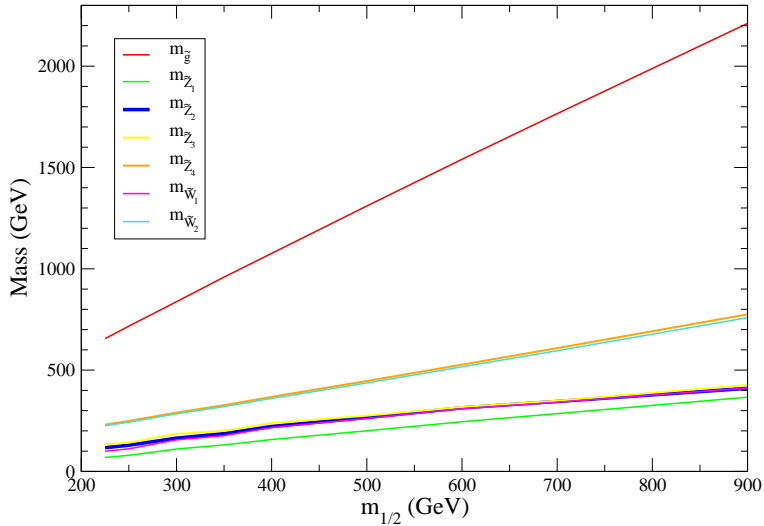}
\caption{Sparticle masses vs. $m_{1/2}$ along lines with constant 
$\Omega_{\tz_1}h^2=0.11$ in the HB/FP region of mSUGRA with $A_0=0$, 
$\tan\beta =30$, $\mu >0$ and $m_t=175$ GeV.}
\label{fig:mass}
\end{center}
\end{figure}

In Fig. \ref{fig:mass}, we show the sparticle mass spectra as a function of $m_{1/2}$
along a line with $\Omega_{\tz_1}h^2\sim 0.11$, for $\tan\beta =30$, $A_0=0$ and $\mu >0$.
The physics in the HB/FP region is not very sensitive to $\tan\beta$ or $A_0$, since the scalar 
masses effectively decouple.
We take $m_t=175$ GeV, but note that the $m_0$ value needed to obtain the correct relic density 
is extremely sensitive to the value of $m_t$ used, as shown in Ref. \cite{bkt}. 
In our case, since the scalar masses are expected
to decouple, the $m_t$ dependence should not matter greatly for the phenomenology of 
interest.\footnote{In our study, we adopt the reference value $m_t=175$ GeV to allow comparisons with other studies.  The recent world average for the $t$-quark mass is $m_t=171.4$ GeV \cite{tmass}.}

While squarks, sleptons and heavy Higgs scalars range in mass 
from $2.5-4.5$ TeV along the range of $m_{1/2}$ shown
in Fig. \ref{fig:mass}, the $\tg$ remains relatively light, of order $650-2200$ GeV.
In addition, since $\mu$ and $m_{1/2}$ are low, the charginos and neutralinos are {\it all}
quite light, and possibly accessible to LHC experiments.
The lower edge of the plot where $m_{1/2}\alt 250$ GeV is excluded by the LEP2 constraint
on the chargino mass.

\begin{figure}[htbp]
\begin{center}
\includegraphics[angle=0,width=0.59\textwidth]{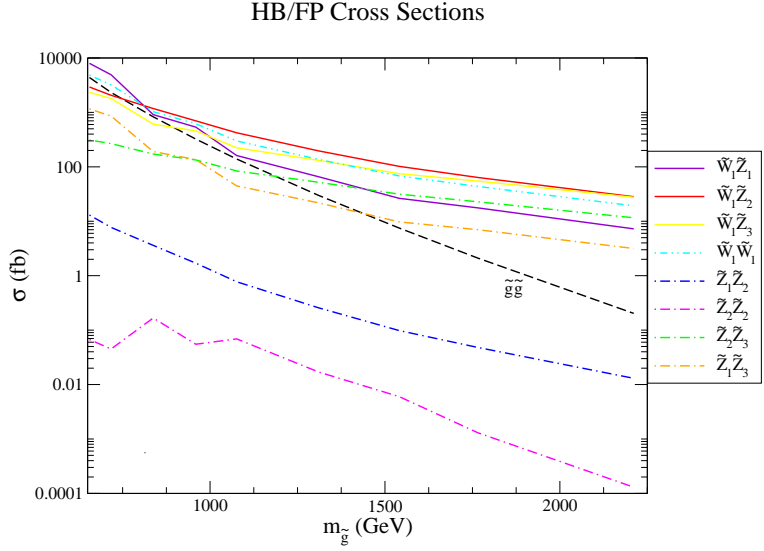}
\caption{Selected sparticle pair production cross sections
vs. $m_{\tg}$ along a line of constant 
$\Omega_{\tz_1}h^2=0.11$ in the HB/FP region of mSUGRA with $A_0=0$, 
$\tan\beta =30$, $\mu >0$ and $m_t=175$ GeV.}
\label{fig:sigma}
\end{center}
\end{figure}

In Fig. \ref{fig:sigma}, we show sparticle pair production rates as a function of $m_{\tg}$
in the HB/FP region. While the production cross sections are evaluated at lowest order in
perturbation theory, we adopt a renormalization/factorization scale choice $Q=(m_1+m_2)/4$ 
for the gluino pair production cross section which gives good agreement 
between LO and NLO results\cite{spira}.\footnote{NLO gluino, chargino and neutralino 
cross sections are shown versus weak scale gaugino mass $M_1$ in the HB/FP region in
Ref. \cite{bkpu}.} For low values of $m_{\tg}\sim 700$ GeV, 
gluino pair production is in the pb range, while a variety of chargino and neutralino 
production processes ({\it e.g.} $\tw_1\tz_{1,2,3}$ and $\tw_1^+\tw_1^-$ production) 
have comparable rates.  For higher values of $m_{\tg}$, the gluino pair production cross section drops quickly,
and is below the fb level for $m_{\tg}>1900$ GeV. The various chargino and neutralino
production rates drop less quickly, and turn out to be by far the dominant 
sparticle production cross sections for $m_{\tg}\agt 1.5$ TeV.

\begin{figure}[htbp]
\begin{center}
\includegraphics[width=0.29\textwidth]{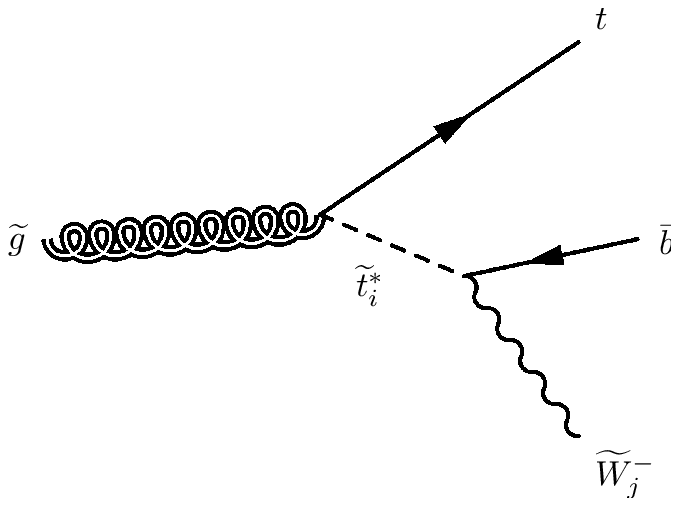}
\hspace{1in}\includegraphics[width=0.29\textwidth]{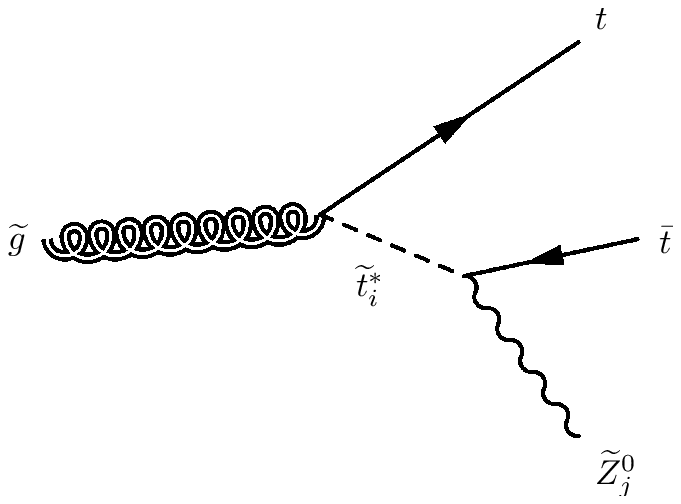}
\caption{Feynman diagrams of dominant gluino decays in the HB/FP region.}
\label{fig:fd}
\end{center}
\end{figure}

In the WMAP-allowed HB/FP region, since squarks have masses in the TeV range, only three-body
decay modes of the gluino are allowed. Moreover, since the $\mu$ parameter is small and the lighter inos
have a large higgsino component, the third generation quark-squark-ino couplings are
enhanced by top quark Yukawa coupling terms\cite{btw,wbsig}, 
and gluinos dominantly decay to third generation particles, especially the top quark. 
Thus, the dominant gluino decays modes in the HB/FP region
consist of $\tg\to t\bar{t}\tz_i$ or $t\bar{b}\tw_j$. Some major $\tg$ branching fractions are listed
in Table \ref{tab:glbf} for a case study which we label as FP5 with $m_{\tg}=1076$ GeV.  The Feynman diagrams of these dominant decay modes are shown in Fig. \ref{fig:fd}.  Thus, we expect in the HB/FP region that $pp\to \tg\tg X$ will yield events with
very large jet and $b$-jet multiplicities, and isolated leptons. However, the combinatoric
backgrounds will likely make kinematic reconstruction of mass edges which depend on $m_{\tg}$
very difficult. Meanwhile, for the same case study as in Table \ref{tab:glbf}, 
since $\tz_2\to e^+e^-\tz_1$ and $\tz_3\to e^+e^-\tz_1$ both occur at a
branching fraction of 3.4\%, it might be possible to see both the $\tz_2-\tz_1$ and
$\tz_3-\tz_1$ mass edges in distributions of invariant opposite-sign/same flavor
isolated dileptons.

\begin{table}
\begin{center}
\begin{tabular}{lc}
\hline
mode & BF \\
\hline
$\tg\to t\bar{t}\tz_1$ & 3.9\%  \\
$\tg\to t\bar{t}\tz_2$ & 14.2\%  \\
$\tg\to t\bar{t}\tz_3$ & 15.0\%  \\
$\tg\to t\bar{t}\tz_4$ & 5.6\%  \\
$\tg\to t\bar{b}\tw_1+c.c$ & 26.8\%  \\
$\tg\to t\bar{b}\tw_2+c.c.$ & 13.9\%  \\
\hline
\end{tabular}
\caption{Selected branching fractions of the $\tg$ for FP5 case study
with parameters $m_0=3050$ GeV, $m_{1/2}=400$ GeV, $A_0=0$, $\tan\beta =30$
and $\mu >0$.
}
\label{tab:glbf}
\end{center}
\end{table}
%

\section{HB/FP signal and background event generation}
\label{sec:evgen}

We use Isajet 7.74\cite{isajet} for the simulation of signal and 
background events at the LHC. A toy detector simulation is employed with
calorimeter cell size
$\Delta\eta\times\Delta\phi=0.05\times 0.05$ and $-5<\eta<5$. The HCAL
energy resolution is taken to be $80\%/\sqrt{E}+3\%$ for $|\eta|<2.6$ and
FCAL is $100\%/\sqrt{E}+5\%$ for $|\eta|>2.6$. 
The ECAL energy resolution
is assumed to be $3\%/\sqrt{E}+0.5\%$. We use a UA1-like jet finding algorithm
with jet cone size $R=0.4$ and require that $E_T(jet)>50$ GeV and
$|\eta (jet)|<3.0$. Leptons are considered
isolated if they have $p_T(e\ or\ \mu)>20$ GeV and $|\eta|<2.5$ with 
visible activity within a cone of $\Delta R<0.2$ of
$\Sigma E_T^{cells}<5$ GeV. The strict isolation criterion helps reduce
multi-lepton backgrounds from heavy quark ($c\bar c$ and $b\bar{b}$) production.

We identify a hadronic cluster with $E_T>50$ GeV and $|\eta(j)|<1.5$
as a $b$-jet if it contains a $B$ hadron with $p_T(B)>15$ GeV and
$|\eta (B)|<3$ within a cone of $\Delta R<0.5$ about the jet axis. We
adopt a $b$-jet tagging efficiency of 60\%, and assume that
light quark and gluon jets can be mis-tagged as $b$-jets with a
probability $1/150$ for $E_T<100$ GeV, $1/50$ for $E_T>250$ GeV, 
with a linear interpolation for $100$ GeV$<E_T<$ 250 GeV. 

We have generated 200K events each for a variety of $m_{1/2}$ values
in the HB/FP region restricted to have $\Omega_{\tz_1}h^2\sim 0.11$.
In addition, we have generated background events using Isajet for
QCD jet production (jet-types include $g$, $u$, $d$, $s$, $c$ and $b$
quarks) over five $p_T$ ranges as shown in Table \ref{tab:bg}. 
Additional jets are generated via parton showering from the initial and final state
hard scattering subprocesses.
We have also generated backgrounds in the $W+jets$, $Z+jets$, 
$t\bar{t}(175)$ and $WW,\ WZ,\ ZZ$ channels at the rates shown in 
Table \ref{tab:bg}. The $W+jets$ and $Z+jets$ backgrounds
use exact matrix elements for one parton emission, but rely on the 
parton shower for subsequent emissions.
\begin{table}
\begin{center}
\begin{tabular}{lccc}
\hline
process & events & $\sigma$ (fb) & cuts C1 (fb)  \\
\hline
QCD ($p_T:50-100$ GeV) & $10^6$ & $2.6\times 10^{10}$ & --\\
QCD ($p_T:100-200$ GeV) & $10^6$ & $1.5\times 10^{9}$ & 1513.3 \\
QCD ($p_T:200-400$ GeV) & $10^6$ & $7.3\times 10^{7}$ & 3873.7 \\
QCD ($p_T:400-1000$ GeV) & $10^6$ & $2.7\times 10^{6}$ & 486.0 \\
QCD ($p_T:1000-2400$ GeV) & $10^6$ & $1.5\times 10^{4}$ & 4.4\\
$W+jets; W\to e,\mu,\tau$ $(p_T(W):100-4000$ GeV)& $5\times 10^5$ & 
$3.9\times 10^{5}$ & 1815.9 \\
$Z+jets; Z\to \tau\bar{\tau},\ \nu s$ $(p_T(Z):100-3000$ GeV) & $5\times 10^5$ & 
$1.4\times 10^{5}$ & 845.3 \\
$t\bar{t}$ & $3\times 10^6$ & $4.6\times 10^{5}$ & 6415.8 \\
$WW,ZZ,WZ$ & $5\times 10^5$ & $8.0\times 10^{4}$ & 9.3 \\
signal (FP5: $m_{\tg}=1076$ GeV) & $2\times 10^5$ & $1.2\times 10^{3}$ & 77.5 \\
\hline
\end{tabular}
\caption{Events generated and cross sections for various SM background 
processes plus one HB/FP case study FP5 
with $m_0=3050$ GeV, $m_{1/2}=400$ GeV, $A_0=0$, $\tan\beta =30$
and $\mu >0$.  The C1 cuts are specified in Eqns. ($4 - 7$).}
\label{tab:bg}
\end{center}
\end{table}

\section{Event characteristics in the HB/FP region}
\label{sec:dist}

We begin by applying a set of pre-cuts to our event samples, which we list 
as cuts set C1\cite{frank}:
\\
\\
\textbf{C1 Cuts:}
\bea
\eslt & >& (100\ {\rm GeV},0.2 M_{eff}),\\
n(jets) &\ge & 4,\\
E_T(j1,j2,j3,j4)& > & 100,\ 50,50,50\ {\rm GeV},\\
S_T &>&0.2 .
\label{c1cutsend}
\eea
Here, $M_{eff}$ is defined as in Hinchliffe {\it et al.}\cite{frank} as
$M_{eff}=\eslt +E_T(j1)+E_T(j2)+E_T(j3)+E_T(j4)$, where $j1-j4$ refer to the
four highest $E_T$ jets ordered from highest to lowest $E_T$, 
$\eslt$ is missing transverse energy and $S_T$ is transverse sphericity.
The event rates in fb are listed after C1 in Table \ref{tab:bg},
and we find that signal with these cuts is swamped by various SM backgrounds, especially those
from QCD multi-jet production and $t\bar{t}$ production.

\begin{figure}[htbp]
\begin{center}
\includegraphics[angle=0,width=0.59\textwidth]{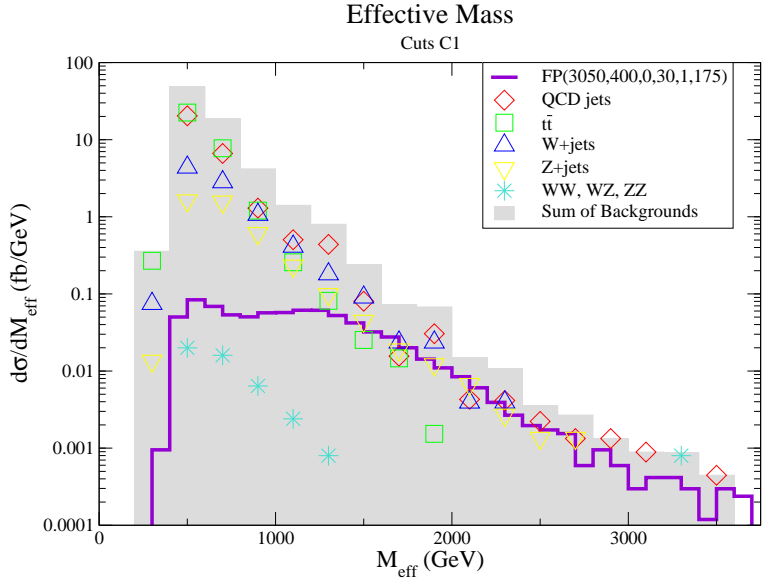}
\caption{Distribution of $M_{eff}$ from the FP5 case study
with $m_0=3050$ GeV, $m_{1/2}=400$ GeV, $A_0=0$, $\tan\beta =30$, $\mu >0$ and $m_t=175$ GeV
(where $m_{\tg}=1076$ GeV), versus various SM backgrounds.}
\label{fig:meff}
\end{center}
\end{figure}

Next, we investigate a variety of distributions. 
We show in Fig. \ref{fig:meff} the $M_{eff}$ distribution after using 
C1. The gray histogram denotes the sum of all backgrounds, while 
individual BG contributions are identified by the legend. The signal for 
case study FP5 is denoted by the purple histogram. In many models 
investigated by Hinchliffe {\it et al.}, it was found that signal emerges
from BG at an $M_{eff}$ value near the peak of the distribution, 
which  in fact provides a rough estimate of the strongly interacting
sparticle masses involved in the production subprocess.
In the HB/FP region, however, squarks have decoupled from the hadronic 
sparticle production cross section, so only gluino pair production contributes.
In addition, since in the HB/FP region gluinos decay via three-body modes,
the average jet $E_T$ is reduced significantly 
compared to SUSY cases with similar sparticle masses but with 
dominantly 2-body decays. Hence, in the HB/FP region, the $M_{eff}$ 
distribution from the signal is typically buried beneath SM BG. In addition, for this case study, we see
some structure to the $M_{eff}$ distribution in the form of 
two separate peaks (which stand out more clearly on a linear scale, when BG is neglected).
The peak near $M_{eff}\sim 500$ GeV comes dominantly from the {\it soft} signal component,
which is mainly high $p_T$ chargino and neutralino production, which after all is the
dominant sparticle production process in the HB/FP region. A second peak around
$M_{eff}\sim 1200$ GeV comes from gluino pair production, which we denote as the
{\it hard} component of the signal. 
%

\begin{figure}[htbp]
\begin{center}
\includegraphics[angle=0,width=0.59\textwidth]{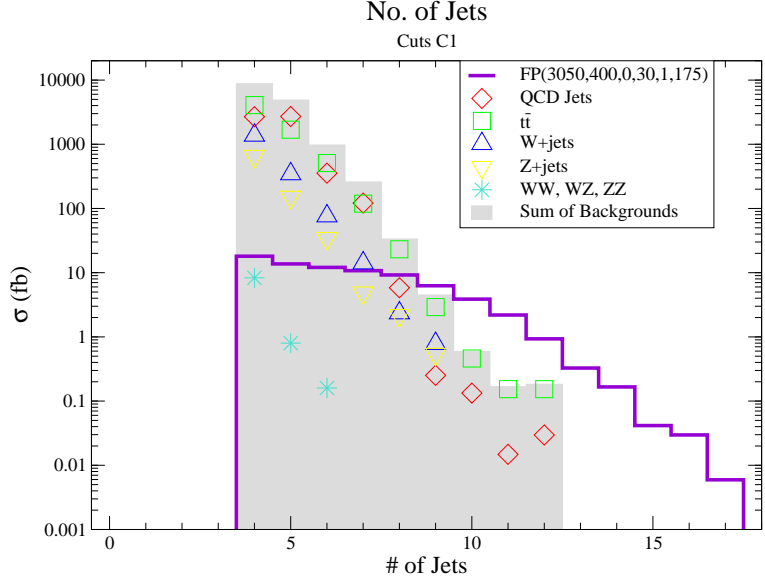}
\caption{Distribution of number of jets in the FP5 case study
with $m_0=3050$ GeV, $m_{1/2}=400$ GeV, $A_0=0$, $\tan\beta =30$, $\mu >0$ and $m_t=175$ GeV
(where $m_{\tg}=1076$ GeV), versus various SM backgrounds.}
\label{fig:njets}
\end{center}
\end{figure}

We noted earlier, based on an examination of gluino decay modes in the HB/FP region, 
that LHC collider events ought to be characterized by large jet multiplicity, 
large $b$-jet multiplicity and large isolated lepton multiplicity.
With this in mind, we show in Fig. \ref{fig:njets} the multiplicity of jets expected
from signal and from SM BG, after cuts C1. At low $n(jets)\sim 4-6$, the 
distribution is dominated by QCD, $t\bar{t}$ and $W,Z+jets$ production. However,
at much higher jet multiplicities $\sim 9-10$, the signal distribution emerges\footnote{The use of the steps in the jet multiplicity was introduced in Ref. \cite{Wnjet} in extracting the signal of top quark pair production.} from the BG.
Of course, at these high jet multiplicities, one may question the validity of the
theoretical BG calculations. However, by investigating QCD multijet production and
$W,Z+jets$ production without imposing C1, it may be possible to normalize
the expected BG distributions to measured data, and thus obtain after LHC turn-on
improved estimates of expected BGs in these channels.

\begin{figure}[htbp]
\begin{center}
\includegraphics[angle=0,width=0.59\textwidth]{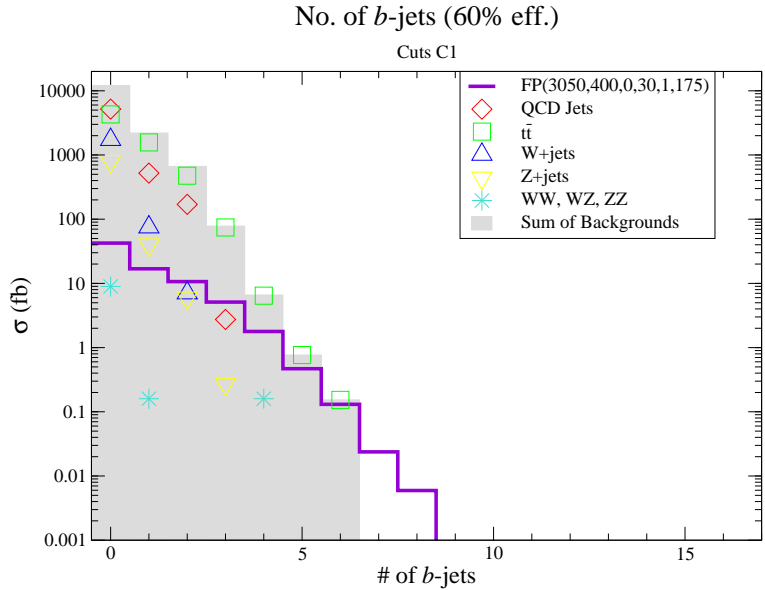}
\caption{Distribution of number of $b$-jets for the FP5 case study
with $m_0=3050$ GeV, $m_{1/2}=400$ GeV, $A_0=0$,$\tan\beta =30$, $\mu >0$ and $m_t=175$ GeV
(where $m_{\tg}=1076$ GeV), versus various SM backgrounds.}
\label{fig:bjets}
\end{center}
\end{figure}

In Fig. \ref{fig:bjets}, we show the expected multiplicity of $b$-jets for signal
and SM BG. The soft component of signal is expected to be $b$-jet poor, since
it comes from hadronic chargino and neutralino decays. However, the hard component
is expected to typically contain at least 4 $b$-jets, aside from efficiency corrections.
Indeed, we see that the signal distribution extends out to high $b$-jet multiplicities
of $n(b-jet)\sim 5-8$, while the  BG typically gives $0-2$ $b$-jets.
As noted previously, Mercadante {\it et al.} exploited this fact to enhance the LHC reach for 
SUSY in the HB/FP region\cite{mmt}.

\begin{figure}[htbp]
\begin{center}
\includegraphics[angle=0,width=0.59\textwidth]{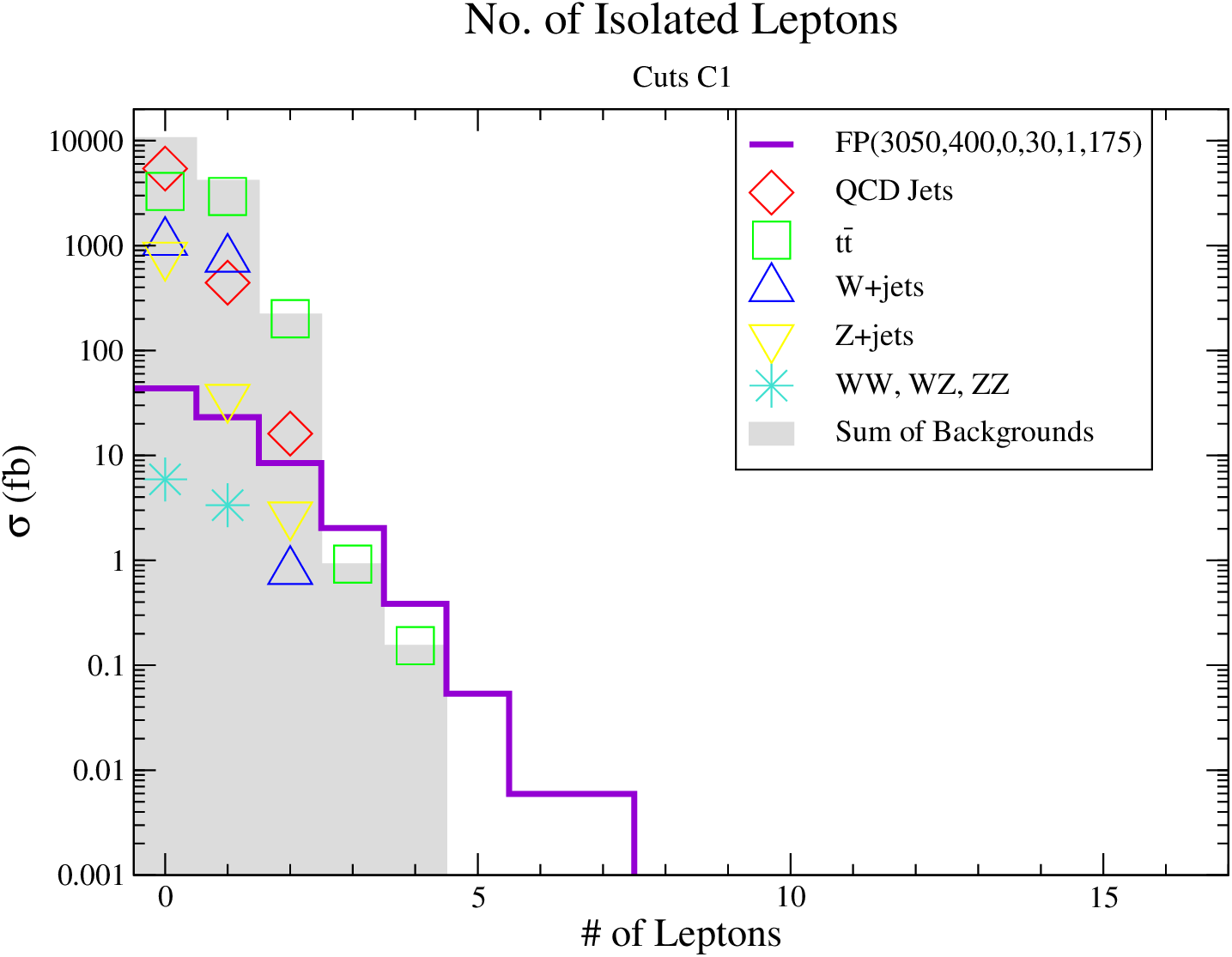}
\caption{Distribution of number of isolated leptons for the FP5 case study
with $m_0=3050$ GeV, $m_{1/2}=400$ GeV, $A_0=0$,$\tan\beta =30$, $\mu >0$ and $m_t=175$ GeV
(where $m_{\tg}=1076$ GeV), versus various SM backgrounds.}
\label{fig:nleps}
\end{center}
\end{figure}

In Fig. \ref{fig:nleps}, we show the multiplicity of isolated leptons: electrons or muons.
Again, while low lepton multiplicity is dominated by SM backgrounds, the high lepton
multiplicity should be dominated by signal, owing to the lengthy gluino cascade decays,
which can spin off additional isolated leptons at various stages.

\begin{figure}[htbp]
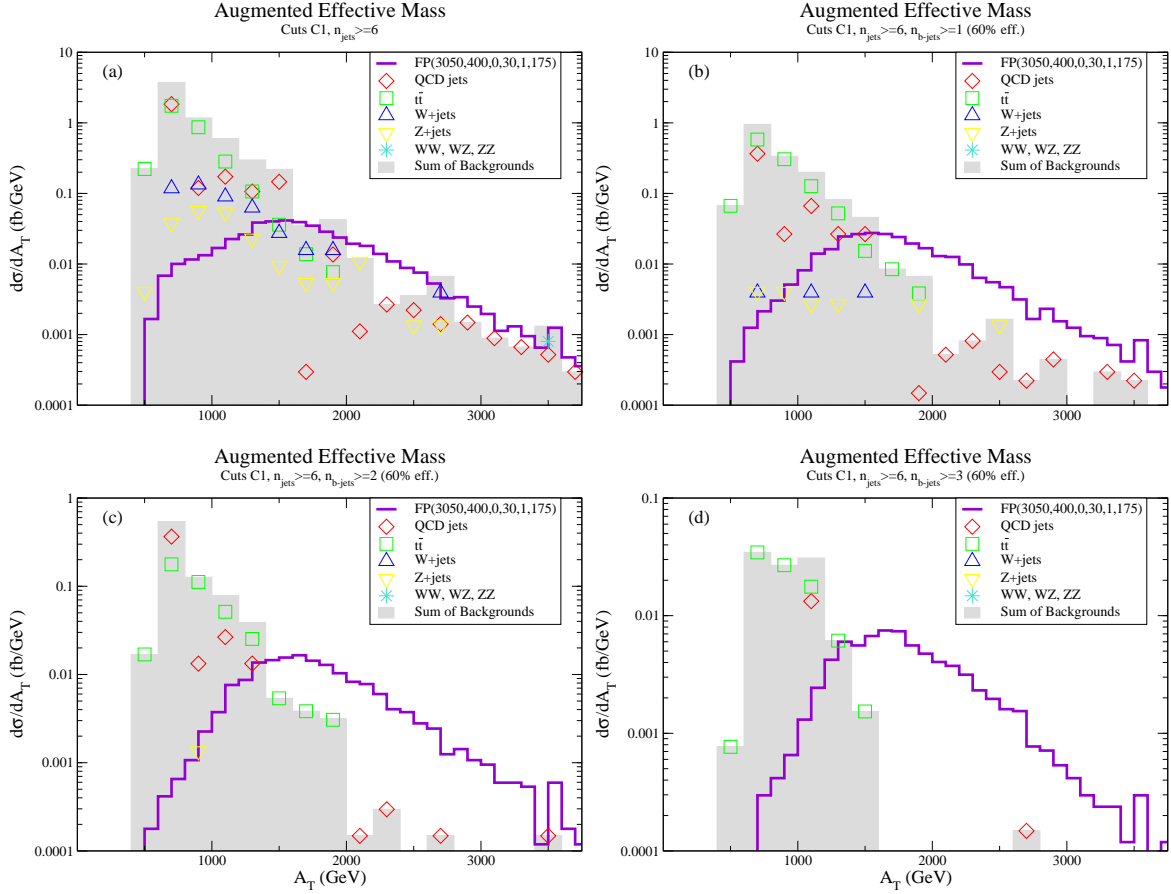

\begin{center}
\includegraphics[angle=0,width=0.45\textwidth]{fp4_cuts1-ameff6j.eps}
\includegraphics[angle=0,width=0.45\textwidth]{fp4+1b_cuts1-ameff6j.eps}\\
\includegraphics[angle=0,width=0.45\textwidth]{fp4+2b_cuts1-ameff6j.eps}
\includegraphics[angle=0,width=0.45\textwidth]{fp4+3b_cuts1-ameff6j.eps}
\caption{Distribution in $A_T$ (defined in Eq. 8) in events with $n(jets)\ge 6$ with varying number of
$b$-tags, for the FP5 case study
with $m_0=3050$ GeV, $m_{1/2}=400$ GeV, $A_0=0$,$\tan\beta =30$, $\mu >0$ and $m_t=175$ GeV
(where $m_{\tg}=1076$ GeV), versus various SM backgrounds.}
\label{fig:meffb}
\end{center}
\end{figure}

At this point, it is evident that requiring collider events with high jet and high $b$-jet 
multiplicity will aid in separating signal from BG in the HB/FP region. Thus, in 
Fig. \ref{fig:meffb}, we show the {\it augmented} effective mass distribution $A_T$,
where
\be
A_T=\eslt +\sum_{leptons}E_T +\sum_{jets}E_T ,
\ee
which gives the added contribution of additional jets beyond $n(jets)=4$ and also 
a contribution from isolated leptons.
The distributions in Fig. \ref{fig:meffb} all contain, along with cuts C1, $n(jets)\ge 6$ and
{\it a}) $n(b-jets)\ge 0$, {\it b}) $n(b-jets)\ge 1$, {\it c}) $n(b-jets)\ge 2$ and 
{\it d}) $n(b-jets)\ge 3$. As we move to higher $b$-jet multiplicity, the signal distribution
begins to stand out clearly from BG, which is dominated at this point by $t\bar{t}$ production. 

\begin{figure}[htbp]
\begin{center}
\includegraphics[angle=0,width=0.59\textwidth]{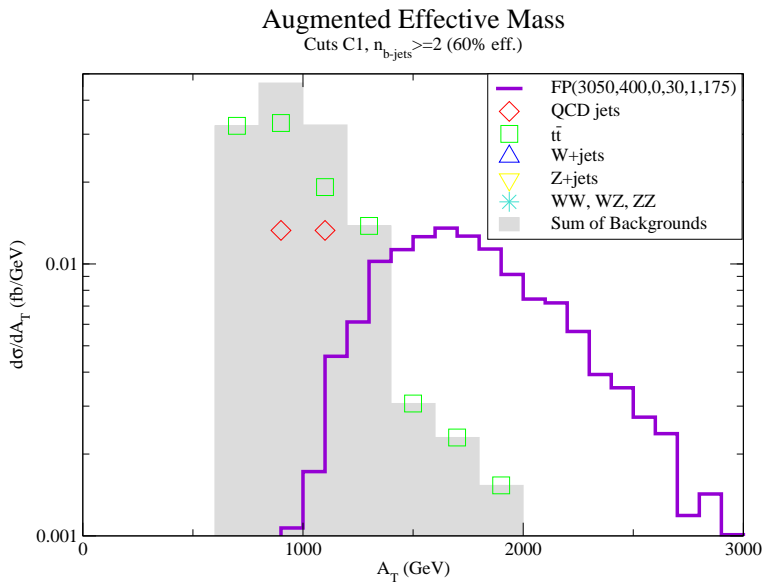}
\caption{Distribution of $A_T$ in events with $\ge 7$ jets and $\ge 2$ $b$-tags, from the FP5 case study
with $m_0=3050$ GeV, $m_{1/2}=400$ GeV, $A_0=0$,$\tan\beta =30$, $\mu >0$ and $m_t=175$ GeV
(where $m_{\tg}=1076$ GeV), versus various SM backgrounds.}
\label{fig:meff7}
\end{center}
\end{figure}

Alternatively, we can move to higher jet multiplicity. In Fig. \ref{fig:meff7}, we
again plot $A_T$ but this time for $n(jets)\ge 7$ and $n(b-jets)\ge 2$. The signal emerges from the BG
clearly above $A_T\sim 1300-1400$ GeV, and has an advantage over Fig. \ref{fig:meffb}{\it d}) in that
a somewhat larger signal rate remains after cuts. For the case shown, by imposing $A_T>1400$ GeV, 
we are left with a signal cross section for case FP5 of 11.1 fb, while BG from $t\bar{t}$ production is at the
$1.5$ fb level with a tiny contribution from QCD multi-jet production. 
In addition, the remaining signal is 98\% from gluino pair production, so is almost entirely
from the hard component of the signal.

\section{Signal, background and sparticle mass extraction}
\label{sec:results}

\begin{figure}[htbp]
\begin{center}
\includegraphics[angle=0,width=0.59\textwidth]{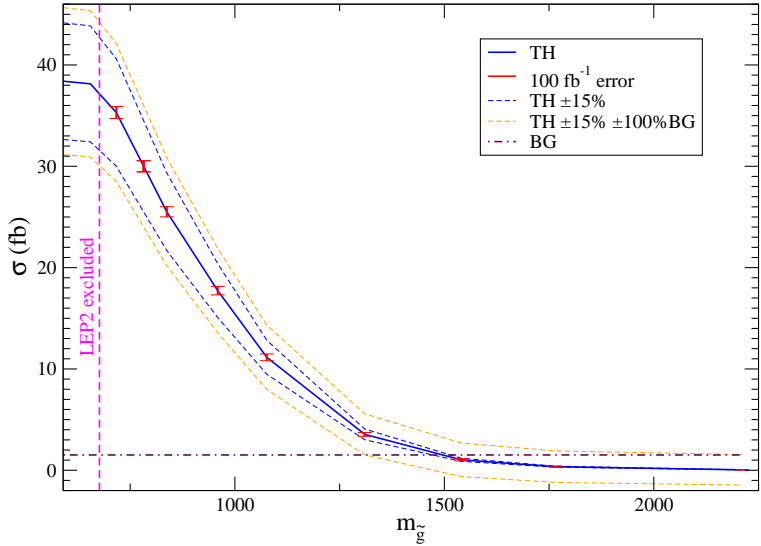}
\caption{Cross section after C1 plus $\ge 7$ jets, $\ge 2$ $b$-tags and
$A_T>1400$ GeV, for various points along the HB/FP region with $\Omega_{\tz_1}h^2\sim 0.11$
with $A_0=0$,$\tan\beta =30$, $\mu >0$ and $m_t=175$ GeV, versus $m_{\tg}$.
We also show a band of the theoretically expected uncertainty of our results due to variations in 
factorization/renormalization scale and variations in $m_{\tq}\sim 2-5$ TeV.
We also show the level of expected SM background.}
\label{fig:sigmgl}
\end{center}
\end{figure}

We will adopt the cuts of Sec. \ref{sec:dist} as our cut set C2:
\\
\\
\textbf{C2 Cuts:}
\beas
& apply\ cut\ set\ C1 & \\
& n(jets)\ge 7 & \\
& n(b-jets)\ge 2 & \\
& A_T\ge 1400\ {\rm GeV}.
\eeas
These cuts have been optimized for $m_{\tg}\sim 1$ TeV. Next, we plot in Fig. \ref{fig:sigmgl} 
the event rate after C2
versus $m_{\tg}$ along a line of FP region with $\Omega_{\tz_1}h^2\sim 0.11$, with 
$\tan\beta =30$, $A_0=0$ and $\mu >0$. For $m_{\tg}\alt 700$ GeV, $m_{\tw_1}< 103.5$ GeV, so the
region is excluded by LEP2 chargino pair searches. The solid blue curve denotes the signal
rate after cuts C2, while the brown dot-dashed curve denotes SM BG. Signal rates are 
typically in the multi-fb regime, and exceed BG out to $m_{\tg}\sim 1500$ GeV. 

Since the signal in Fig. \ref{fig:sigmgl} comes from nearly pure $\tg\tg$ production, 
the total rate can be used as an absolute measure of the gluino mass. There are of course a variety of 
theoretical uncertainties which arise. One comes from how well-known is the absolute gluino pair
production cross section. The value of $\sigma (\tg\tg )$ has been computed to NLO in QCD in
Ref. \cite{spira}, where it is shown that a variation in renormalization/factorization 
scale leads to an uncertainty in $\sigma (\tg\tg )$ of $\pm 11\%$. A further uncertainty arises from
variations in the squark mass. Here, we are assuming decoupled scalars, so variation due to changes
in $m_{\tq}$ are expected to be small. Nonethless, we find that by varying $m_{\tq}:2-5$ TeV,
the cross section still varies by $\pm 10\%$. Folding the NLO uncertainty  in quadrature with the 
$m_{\tq}$ uncertainty, we estimate the cross section uncertainty at $\pm 15\%$, and plot the 
expected theory cross section variation as the blue dashed lines.

\begin{figure}[htbp]
\begin{center}
\includegraphics[angle=-90,width=0.49\textwidth]{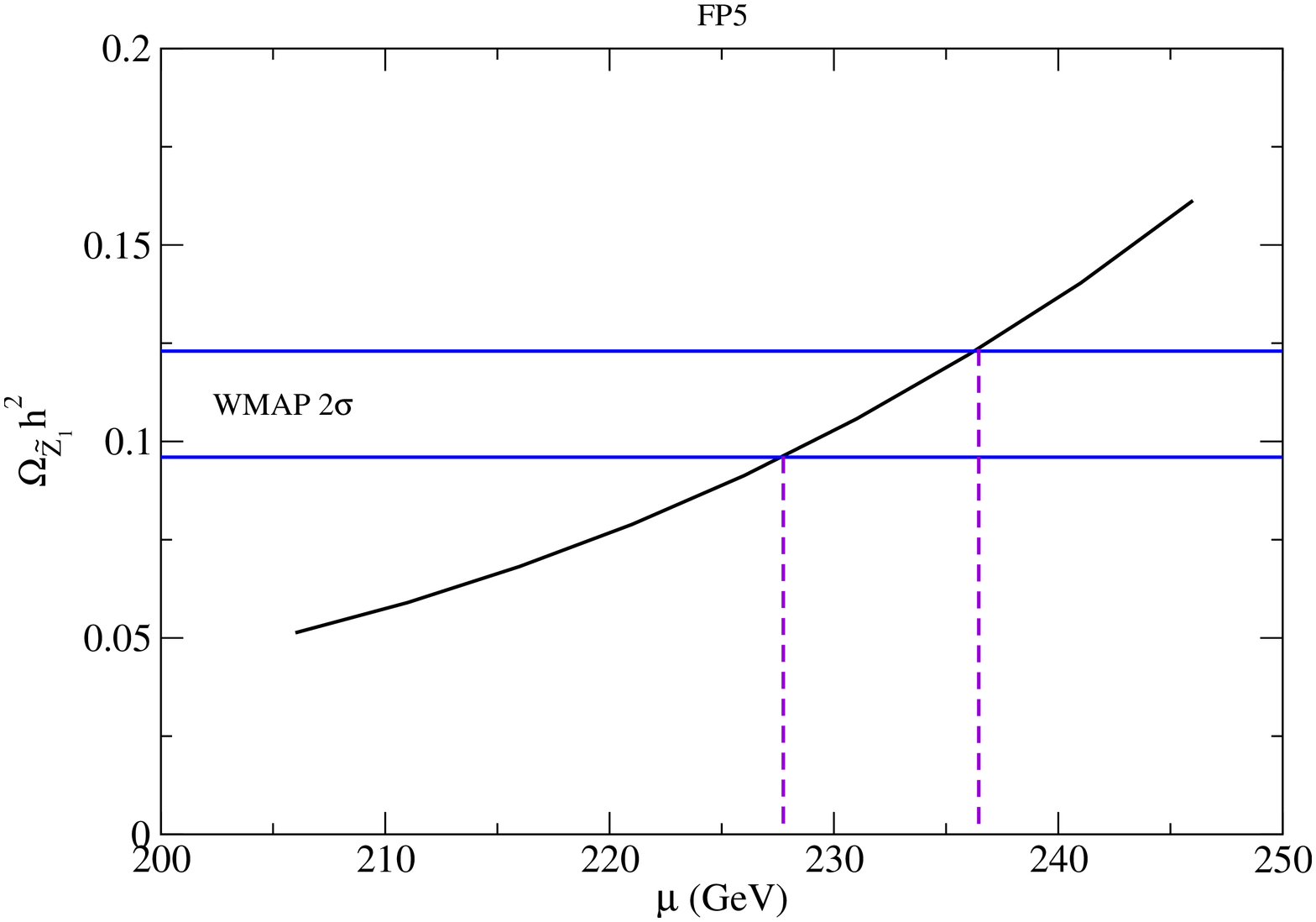}\\
\vspace{.2in}\includegraphics[angle=-90,width=0.49\textwidth]{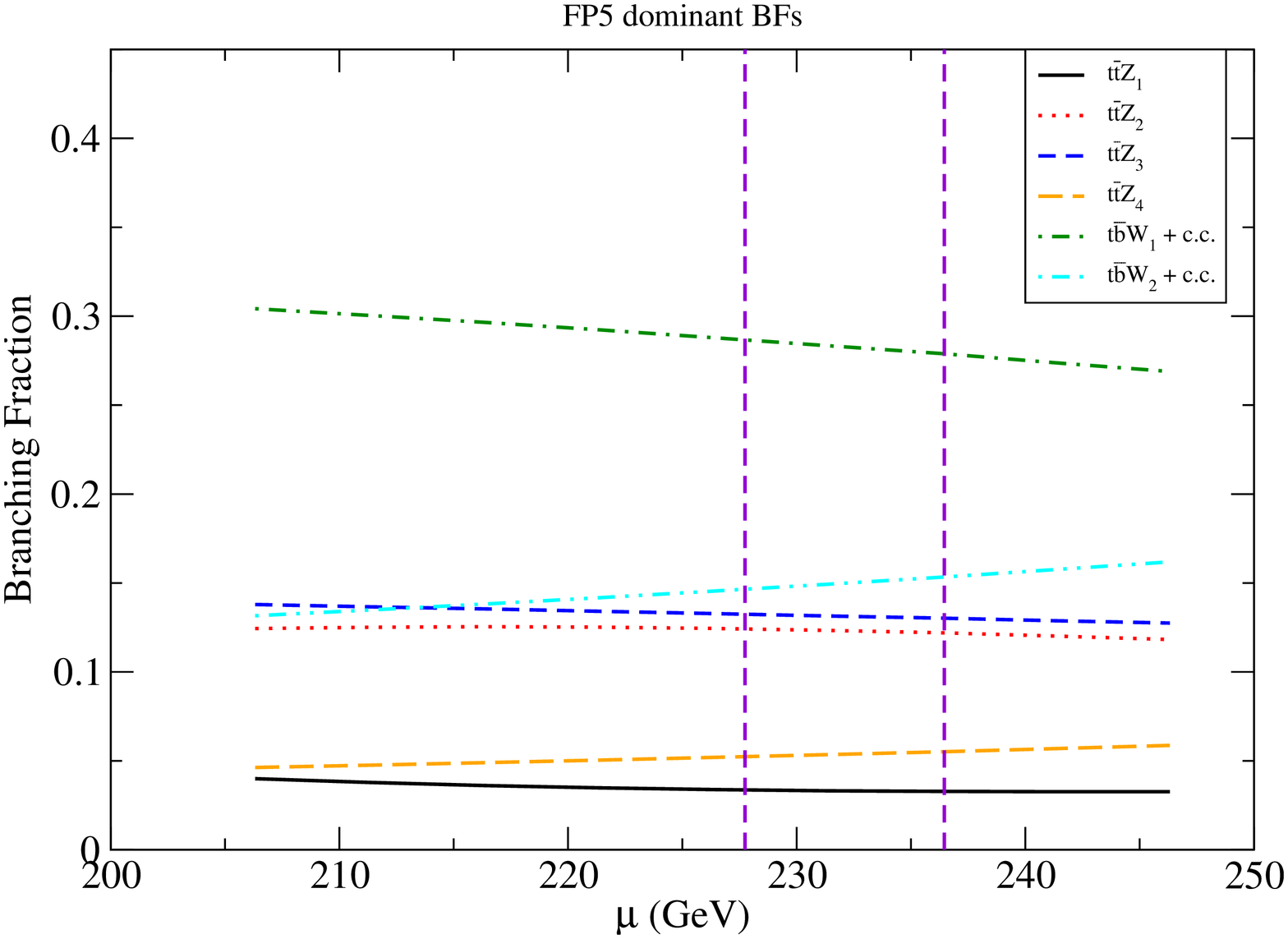}
\caption{{\it a}) Variation in neutralino relic density with variation in $\mu$ for case study FP5.
In {\it b}), we show variation in dominat gluino branching fractions versus $\mu$.}
\label{fig:bfs}
\end{center}
\end{figure}

At this point, it can be asked how well will we know the gluino branching fractions, upon 
which the signal rate also depends. Here, we remark that in the region with decoupled scalars, we are 
relying on a value of $\mu$ that is just right so that the neutralino LSP saturates the CDM relic density
measurement. Small variations in $\mu$ about this region are found to lead to 
only small changes in the gluino branching fractions. This is shown in Fig. \ref{fig:bfs},
where we plot in frame {\it a}) variations in $\Omega_{\tz_1}h^2$ versus $\mu$, and
in frame {\it b}) variations in the dominant gluino branching fractions. 
In the plot, we adopt as usual the case study FP5, and vary $\mu$ by adopting the 
non-universal Higgs soft mass model\cite{nuhm} in Isajet, which allows use of mSUGRA parameters, but also
independent variation in the $\mu$ and $m_A$ parameters (we keep $m_A$ fixed).

It might also be argued that the event rate depends on the value of $\tan\beta$ that we have 
selected for our case study. In fact, since scalar masses have decoupled, $b$ and $\tau$ Yukawa 
coupling effects are tiny, and the variation of the signal after cuts C2 with $\tan\beta$ 
is comparatively negligible, as long as we require that the $\mu$ value be fixed so that one
obtains the relic density $\Omega_{\tz_1}h^2\sim 0.11$. This is illustrated in Table \ref{tab:tanb},
where we plot signal rate after cuts C2 for $\tan\beta =10$, 20, 30, 40 and 50. In each case, 
the value of $m_{1/2}$ is fixed at 400 GeV, but $m_0$ is chosen so that the correct relic density
is obtained. The resulting cross section after cuts C2 shows only a $\pm 6\%$ variability.
Meanwhile, variations in the $A_0$ parameter again mainly affect the scalar sector, but since these
decouple, the effects should again be small.
%
\begin{table}
\begin{center}
\begin{tabular}{lcc}
\hline
$m_0$ & $\tan\beta$ & $\sigma $(C2) (fb)  \\
\hline
4090 & 10 & 9.92 \\
3150 & 20 & 10.45 \\
3050 & 30 & 11.15 \\
3000 & 40 & 11.04 \\
2970 & 50 & 11.17 \\
\hline
\end{tabular}
\caption{Cross section after cuts C2 for HB/FP cases with $m_{1/2}=400$ GeV,
$A_0=0$, $\mu >0$ and  $m_t=175$ GeV.
We list the $m_0$ value required to give $\Omega_{\tz_1}h^2\sim 0.11$
for different $\tan\beta$ values.
}
\label{tab:tanb}
\end{center}
\end{table}

A further consideration is to ask how well we really know our background estimates. 
At this stage, the answer is difficult to know, and depends on several factors, including
how well the selected event generator, Isajet, models SM backgrounds. If indeed $t\bar{t}$
production is the dominant BG, then the plethora of $t\bar{t}$ events 
produced at the LHC will allow detailed study of this reaction, so that the distributions will be
well-known from data. 
Better theory modeling-- such as inclusion of exact matrix elements for extra 
jet radiation\cite{bcr}-- will also help.
Likewise, it can be expected that $W+jets$, $Z+jets$ and QCD backgrounds
will also be well-studied, and the high $n(jet)$ and high $A_T$ tails will be better known due to
actual collider measurements. 
In any case, we try to make a rough estimate by simply assuming that our event generator
background is known to $\pm 100\%$. We add and subtract this BG uncertainty to our theory curves in Fig. \ref{fig:sigmgl}, with the resultant band being denoted by orange dashed lines.

At this point, we can try to estimate the precision with which the gluino mass can be extracted 
from total cross section measurements. We show in Fig. \ref{fig:sigmgl} as data points the 
error bars expected from measuring the total cross section after cuts C2 with an assumed 100 fb$^{-1}$ of
integrated luminosity (red data points). 
A simple estimate of the uncertainty can be gained from the intersection of the upper and lower limits
on the statistical cross section measurement with the band of theory uncertainty. Using this method,
we find that points 2-6 yield a gluino mass measured in the range of $\pm 8\%$, as shown in 
Table \ref{tab:fppoints}. 
The precision will increase or decrease
depending on the ultimate uncertainty ascribed to the BG by the experimental groups. 
It would also decrease if an NNLO computation of gluino pair production is made.
Note that even 
if the statistical error bars drop to zero (infinite integrated luminosity), the theory uncertainty
still gives $\sim 7$\% uncertainty.
FP1-- which is below the LEP2 excluded boundary-- is difficult to measure because the projected
theory curves level off for lower values of $m_{\tg}$. This is just a result of the
fact that we optimized cuts in the 1 TeV $m_{\tg}$ region. A better optimization with softer cuts 
would need to be performed to extract these lower gluino masses. For $m_{\tg}\agt 1300$ GeV,
another optimization would be needed with harder cuts. Here, the absolute gluino pair event rate 
is dropping, so we expect a rate-based measurement of $m_{\tg}$ would be more challenging and perhaps 
not feasible in this higher mass region.

\section{Leptonic signatures}
\label{sec:leps}

\begin{figure}[htbp]
\begin{center}
\includegraphics[angle=0,width=0.59\textwidth]{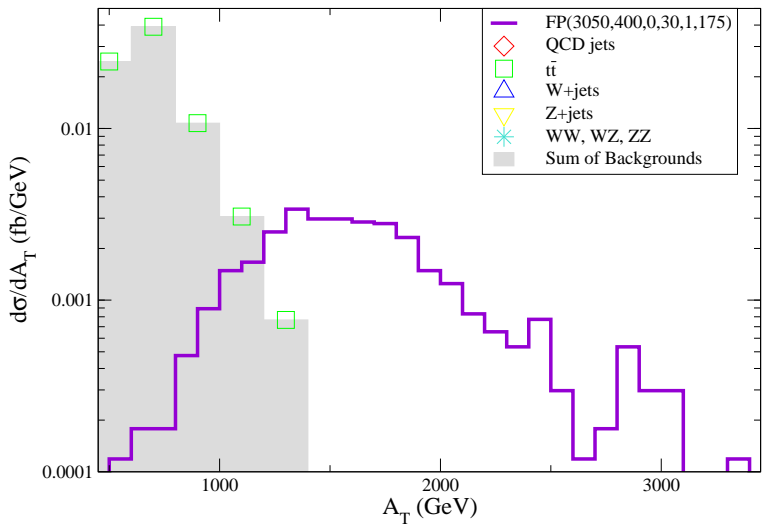}
\caption{Distribution in $A_T$ of events with cuts C1, $n(leps)\ge 2$, $n(b-jets)\ge 2$ and
$n(jets)\ge 4$ for the FP5 case study with $m_{\tg}=1076$ GeV.} 
\label{fig:ATlepsge2}
\end{center}
\end{figure}

While the analysis in Sec. \ref{sec:results} focussed on lepton-inclusive signals, it is also
useful to make use of the isolated lepton content of the signal. We expect events 
containing multiple isolated leptons to have somewhat reduced jet multiplicity compared to
events with zero or one isolated lepton. To proceed with the multi-lepton channels, 
we retained cuts C1 and examined the $A_T$ distribution maintaining $n(b-jets)\ge 2$ but
requiring $n(jets)\ge 4$ or $5$. The distribution in $A_T$ for $n(jets)\ge 4$ is shown in 
Fig. \ref{fig:ATlepsge2}. Here we see signal emerging from BG for $A_T>1200$ GeV.
(The plot using $n(jets)\ge 5$ is similar, but with lower signal and BG rates.)
Hence, we adopt cut set C3 for events with 2 or more isolated leptons:
\\
\\
\textbf{C3 Cuts:}
\beas
& cuts\ set\ C1 & \\
& n(isol.\ leptons)\ge 2 & \\
& n(jets)\ge 4 & \\
& b(b-jets)\ge 2 & \\
& A_T\ge 1200\ {\rm GeV} 
\eeas

\begin{figure}[htbp]
\begin{center}
\includegraphics[angle=0,width=0.59\textwidth]{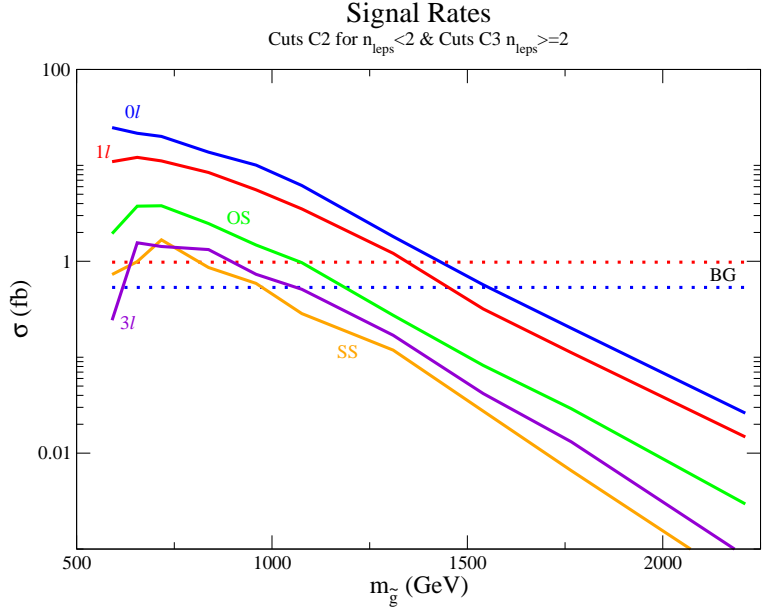}
\caption{Event rates for zero and one isolated lepton  events after cuts C2, 
and OS and SS dileptons and trileptons after cuts C3, versus 
$m_{\tg}$.  Zero and one lepton backgrounds are shown with the same color as the signal.  We found no backgrounds events to OS, SS and $3l$ to the cross section level shown.}
\label{fig:sigleps}
\end{center}
\end{figure}

In Fig. \ref{fig:sigleps}, we show the signal rate of various multi-lepton topologies
versus $m_{\tg}$ for FP cases with $A_0=0$, $\tan\beta =30$ and $\mu >0$.
The zero and one lepton topologies use cuts C2, while the same-sign (SS) dilepton, 
opposite sign dilepton (OS) and trilepton rates use cuts C3. We see that there should be consistent
signals above SM backgrounds in all the various multi-lepton channels for much of the mass
range of $m_{\tg}$.  Same sign lepton events would establish the Majorana nature of the gluino\cite{ssdl}.

\begin{figure}[htbp]
\begin{center}
\includegraphics[angle=0,width=0.59\textwidth]{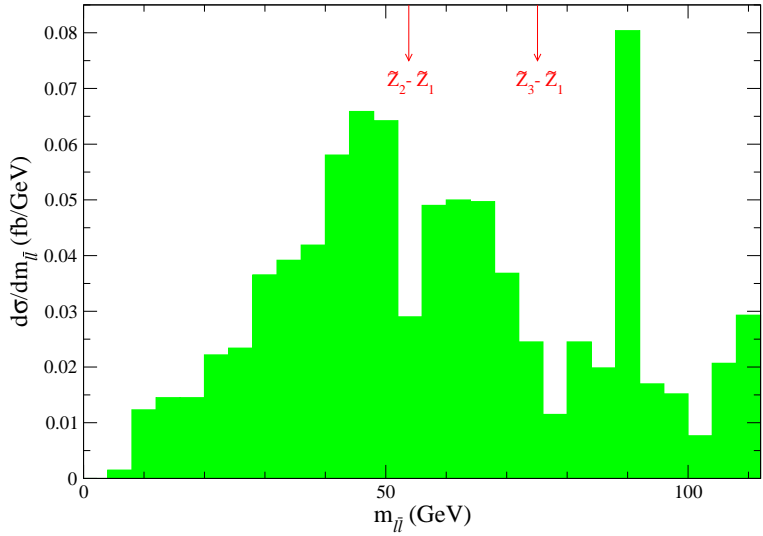}
\caption{Distribution in OS/SF dilepton invariant mass for 
case FP4 using cuts C3. Two mass edges are becoming apparent at a luminosity of 100 fb$^{-1}$, in addition to the $Z$ peak.}
\label{fig:mll}
\end{center}
\end{figure}

It is also well-known that kinematic information on neutralino mass differences can be gleaned
by examining the invariant mass distribution of opposite-sign/same flavor 
dilepton pairs (OS/SF)\cite{mll}.
We plot in Fig. \ref{fig:mll} the invariant mass distribution for case study  FP4.
The HB/FP region is characterized by the fact that $m_{\tz_2}-m_{\tz_1}< M_Z$ {\it and} by
$m_{\tz_3}-m_{\tz_1}<M_Z$, so that two-body spoiler decays of $\tz_2$ and $\tz_3$ are closed. We then
expect two mass edges in the $m(\ell^+\ell^- )$ distribution: in the case of FP4, 
one is at  $m_{\tz_2}-m_{\tz_1}=53.8$ GeV
and another at $m_{\tz_3}-m_{\tz_1}=75.1$ GeV. Indeed, the double mass edge structure 
is becoming visible in the $M_{l \bar l}$ distribution with 100 fb$^{-1}$ of data as shown in Fig. \ref{fig:mll}.
%

\section{Conclusions}
\label{sec:conclude}

In this paper, we have examined the sort of collider events to be expected at the CERN LHC for
SUSY models in the HB/FP region of the mSUGRA model. 
We found that by requiring high jet and $b$-jet multiplicity, and a high effective mass cut,
a rather pure signal emerged from a dominantly $t\bar{t}$ SM background. Since the signal came almost entirely
from gluino pair production, and the decay branching fractions were fixed by assuming the 
neutralino relic density saturated the WMAP $\Omega_{\tz_1}h^2$ measurement, 
the total signal rate could be used to extract an estimate of the gluino
mass. Factoring in theory uncertainty on the total cross section and a $\pm 100\%$ error 
estimate on remaining background, we found that $m_{\tg}$ could be measured to a precision of about 8\%
for 100 fb$^{-1}$ of integrated luminosity. This was our central result.

We note here that our conclusions apply more generally than to just the HB/FP region of the mSUGRA model.
The key assumptions needed for our analysis are that:
\begin{enumerate}
\item The flavor/CP conserving MSSM is the correct effective theory of nature at the weak scale, with the lightest
neutralino as LSP.
\item We assume gaugino mass unification, as occurs in many SUSY GUT and string models.
\item We assume that scalars-- the squarks and sleptons-- decouple due to mass values beyond the few TeV level.
This leaves just the various gluinos, charginos and neutralinos contributing to LHC collider events.
\item The value of $\mu$ is fixed by the requirement that the relic abundance of $\tz_1$ saturates
the WMAP measured value. This, along with gaugino mass unification,  
fixes the sparticle branching fractions to their assumed values.
\end{enumerate}
If these conditions are fulfilled, then the methods presented here should allow for a gluino mass
extraction if $m_{\tg}$ is in the mass range of $\sim 700-1300$ GeV. 
We note here that our result depends on the sparticle branching fractions being fixed to
values near our calculated results. These in turn depend on the above assumptions being fulfilled.
Thus, our study should be
applicable to other heavy scalar situations, not only the HB/FP region of mSUGRA. Recently, such
models have received renewed attention in light of FCNC
constraints, and some string theory motivated models have produced heavier
scalar spectra\cite{string}. Our considerations also apply to the low scalar mass regime of 
split SUSY models\cite{splitss},
where the gluino decays promply inside collider detectors.

In addition, we note that the signal can be separated as to its isolated lepton content.
Typically, for each additional isolated lepton, there should be on average 1.5 less jets per event.
The OS/SF dilepton mass distribution embedded in the hard signal component should exhibit
mass edges at $m_{\tz_2}-m_{\tz_1}$ and also at $m_{\tz_3}-m_{\tz_1}$, which are distinctive of this
scenario in which the LSP is a mixed bino-higgsino particle. The same mass edges should appear in
the clean trilepton channel originating mainly from chargino-neutralino production 
(the soft component), as shown in Ref. \cite{bkpu}.
The mass-difference edges, along with the absolute gluino mass, may provide enough information to
constrain the absolute chargino and neutralino masses (including the LSP mass), under the assumptions listed above.  

\section*{Acknowledgments}
This work was supported in part by the U.S.~Department of Energy under grant No. DE-FG02-95ER40896, 
by the Wisconsin Alumni Research Foundation.  We thank T. Krupovnickas and X. Tata for discussions.
The work of L.W.  is supported by the National Science Foundation under Grant
No. 0243680 and the Department of Energy under grant \# DE-FG02-90ER40542.
Any opinions, findings, and conclusions or recommendations expressed in this
material are those of the author(s) and do not necessarily reflect the views
of the National Science Foundation.


\begin{thebibliography}{99}
\small
%
%
\bibitem{sugra} E. Cremmer, S. Ferrara, L. Girardello and A. van Proeyen,
\npb{212}{1983}{413}.
%
\bibitem{msugra} A.~Chamseddine, R.~Arnowitt and P.~Nath, 
\prl{49}{1982}{970};
R.~Barbieri, S.~Ferrara and C.~Savoy, 
\plb{119}{1982}{343};
N.~Ohta, \ptp{70}{1983}{542};
L.~J.~Hall, J.~Lykken and S.~Weinberg, \prd{27}{1983}{2359};
for reviews, see H.~P.~Nilles, {\em Phys.~Rep.} {\bf 110} (1984) 1, and
P.~Nath, \hepph{0307123}.
%
\bibitem{wss} For an overview, see {\it e.g.} 
H.~Baer and X.~Tata, {\em Weak Scale Supersymmetry}, 
Cambridge University Press (2006).
%
\bibitem{isajet} ISAJET v7.74, by H.~Baer, F.~Paige, S.~Protopopescu and
X.~Tata, \hepph{0312045}.
%
\bibitem{kraml} H.~Baer, J.~Ferrandis, S.~Kraml and W.~Porod, 
\prd{73}{2006}{015010}.
%
\bibitem{bbb} H.~Baer, C.~Balazs and A.~Belyaev, \jhep{0203}{2002}{042}
%
\bibitem{wmap} D. N. Spergel {\it et al.} (WMAP Collaboration),
\astroph{0603449} (2006).
%
\bibitem{bulk} H.~Baer and M.~Brhlik, \prd{53}{1996}{597};
V.~Barger and C.~Kao, \prd{57}{1998}{3131}.
%
\bibitem{bb3} J.~Ellis, K.~Olive, Y.~Santoso and V.~Spanos,
\plb{565}{2003}{176};
H.~Baer and C.~Balazs, JCAP{\bf 05} (2003) 006;
U.~Chattapadhyay, A.~Corsetti and P.~Nath, \prd{68}{2003}{035005};
A.~Lahanas and D.~V.~Nanopoulos, \plb{568}{2003}{55};
A.~Djouadi, M.~Drees and J.~Kneur, \hepph{0602001}
%
\bibitem{stau} J. Ellis, T. Falk and K. Olive,
\plb{444}{1998}{367}; J. Ellis, T. Falk, K. Olive and M. Srednicki,
\app{13}{2000}{181};
M.E. G\'{o}mez, G. Lazarides and C. Pallis, \prd{61}{2000}{123512}
and \plb{487}{2000}{313};
A. Lahanas, D. V. Nanopoulos and V. Spanos, \prd{62}{2000}{023515};
R.~Arnowitt, B.~Dutta and Y.~Santoso,
\npb{606}{2001}{59};
H. Baer, C. Balazs and A. Belyaev, \jhep{0203}{2002}{042}.
%
\bibitem{stop} C.~B\"ohm, A.~Djouadi and M.~Drees,
  \prd{30}{2000}{035012};
J.~R.~Ellis, K.~A.~Olive and Y.~Santoso, \app{18}{2003}{395};
J.~Edsj\"o {\it et al.}, JCAP {\bf 04} (2003) 001
%
\bibitem{Afunnel} M. Drees and M. Nojiri, \prd{47}{1993}{376};
H. Baer and M. Brhlik, \prd{53}{1996}{597} and \prd{57}{1998}{567};
H. Baer, M. Brhlik, M. Diaz, J. Ferrandis, P. Mercadante,
P. Quintana and X. Tata, \prd{63}{2001}{015007};
J. Ellis, T. Falk, G. Ganis, K. Olive and M. Srednicki,
\plb{510}{2001}{236}; L. Roszkowski, R. Ruiz de Austri and T. Nihei,
\jhep{0108}{024}{2001}; A. Djouadi, M. Drees and J. L. Kneur,
\jhep{0108}{2001}{055};
A. Lahanas and V. Spanos, \epjc{23}{2002}{185}.
%
\bibitem{drees_h} R.~Arnowitt and P.~Nath, \prl{70}{1993}{3696};
H.~Baer and M.~Brhlik, Ref.~\cite{bulk};
A.~Djouadi, M.~Drees and J.~Kneur, \plb{624}{2005}{60}.
%
\bibitem{ccn} K.~L.~Chan, U.~Chattopadhyay and P.~Nath, \prd{58}{1998}{096004}.
%
\bibitem{fmm} J.~Feng, K.~Matchev and T.~Moroi, \prl{84}{2000}{2322}, 
\prd{61}{2000}{075005} and \prd{63}{2001}{095003}; 
J. Feng and F. Wilczek, \plb{631}{2005}{170}.
%
\bibitem{hb_fp} The HB/FP region appears much earlier in 
H.~Baer, C.~H.~Chen, F.~Paige and X.~Tata, \prd{52}{1995}{2746} and 
\prd{53}{1996}{6241}, but is not named, and fine-tuning is not addressed.
%
\bibitem{lhcreach} H.~Baer, C.~H.~Chen, F.~Paige and X.~Tata, 
Ref. \cite{hb_fp}; H.~Baer, C.~H.~Chen, M.~Drees, F.~Paige and X.~Tata, 
\prd{59}{1999}{055014}; S.~Abdullin and F.~Charles, \npb{547}{1999}{60};
S.~Abdullin {\it et al.} (CMS Collaboration), \hepph{9806366};
B.~Allanach, J.~Hetherington, A.~Parker and B.~Webber, 
\jhep{08}{2000}{017}.
%
\bibitem{bbbkt} H.~Baer, C.~Balazs, A.~Belyaev, T.~Krupovnickas and X.~Tata,
\jhep{0306}{2003}{054}.
%
\bibitem{bbkt} H. Baer, A. Belyaev, T. Krupovnickas and X. Tata,
\jhep{0402}{2004}{007}; H. Baer, T. Krupovnickas and X. Tata,
\jhep{0406}{2004}{061}.
%
\bibitem{trilep} H. Baer and X. Tata, \prd{47}{1993}{2739}.
%
\bibitem{Barger:1998hp}
V.~D.~Barger and C.~Kao, \prd{60}{1999}{115015}, \hepph{9811489}.
H. Baer, M. Drees, F. Paige, P. Quintana and X. Tata,
\prd{61}{2000}{095007};
K. Matchev and D. Pierce, \plb{467}{1999}{225}.
%
\bibitem{frank} I. Hinchliffe, F. Paige, M. Shapiro, 
J. S\"oderqvist and W. Yao,
\prd{55}{1997}{5520}.
%
\bibitem{bkt} H. Baer, T. Krupovnickas and X. Tata,
\jhep{0307}{2003}{020}.
%
\bibitem{tmass}
Brubaker, E. {\it et al.} \hepex{0608032}

\bibitem{direct} For a recent analysis, see H.~Baer, C.~Balazs,
A.~Belyaev and J.~O'Farrill, JCAP{\bf 0309}, 2003 (007); a subset of
earlier work includes M.~Goodman and E.~Witten, \prd{31}{1985}{3059};
K.~Griest, \prl{61}{1988}{666} and \prd{38}{1988}{2357} [Erratum-ibid.\
D {\bf 39}, 3802 (1989)]; M.~Drees and M.~Nojiri, \prd{47}{1993}{4226}
and \prd{48}{1993}{3483}; V.~A.~Bednyakov, H.~V.~Klapdor-Kleingrothaus
and S.~Kovalenko, \prd{50}{1994}{7128}; P.~Nath and R.~Arnowitt,
\prl{74}{1995}{4592}; L.~Bergstrom and P.~Gondolo, \app{5}{1996}{263};
H.~Baer and M.~Brhlik, \prd{57}{1998}{567}; J.~Ellis, A.~Ferstl and
K.~Olive, \plb{481}{2000}{304} and \prd{63}{2001}{065016}; E.~Accomando,
R.~Arnowitt, B.~Dutta and Y.~Santoso, \npb{585}{2000}{124}; A.~Bottino,
F.~Donato, N.~Fornengo and S.~Scopel, \prd{63}{2001}{125003};
M.~E.~Gomez and J.~D.~Vergados, \plb{512}{2001}{252}; A.~B.~Lahanas,
D.~V.~Nanopoulos and V.~C.~Spanos, \plb{518}{2001}{94}; A.~Corsetti and
P.~Nath, \prd{64}{2001}{115009}; E.~A.~Baltz and P.~Gondolo,
\prl{86}{2001}{5004}; M.~Drees, Y.~G.~Kim, T.~Kobayashi and
M.~M.~Nojiri, \prd{63}{2001}{115009}; see also J.~Feng, K.~Matchev and
F.~Wilczek, \plb{482}{2000}{388} and \prd{63}{2001}{045024}; R.~Ellis,
A.~Ferstl, K.~A.~Olive and Y.~Santoso, \prd{67}{2003}{123502};
J.~R.~Ellis, K.~A.~Olive, Y.~Santoso and V.~C.~Spanos,
\prd{69}{2004}{015005}; see C.~Mu\~noz, \hepph{0309346} for a recent
review.
%
\bibitem{indirect} 
J. Feng, K. Matchev and F. Wilczek, \plb{482}{2000}{388} 
and \prd{63}{2001}{045024};
H. Baer, A. Belyaev, T. Krupovnickas and J. O'Farrill,
\jhep{0408}{2004}{005}.
%
\bibitem{hp} I. Hinchliffe and F. Paige, 
in {\it Workshop on Physics at TeV Colliders, Les Houches, France, 21 May - 1 Jun 2001}.
%
\bibitem{meff-hq} H.~Baer, V.~D.~Barger and R.~J.~N.~Phillips, \prd{39}{1989}{3310}
%
\bibitem{mmt} P.~G.~Mercadante, J.~K.~Mizukoshi and X.~Tata,
\prd{72}{2005}{035009}.
%
\bibitem{bkpu} H.~Baer, T.~Krupovnickas, S.~Profumo and P.~Ullio, 
\jhep{0510}{2005}{020}.
%
\bibitem{baltz} E. Baltz, M. Battaglia, M. Peskin and T. Wizansky, 
\prd{74}{2006}{103521}.
%
\bibitem{sscgaugino} H.~Baer, V.~D.~Barger, D.~Karatas and X.~Tata,
\prd{35}{1987}{96}.
%
\bibitem{glmass}
H.~Bachacou, I.~Hinchliffe and F.~E.~Paige, \prd{62}{2000}{015009}.
B.~C.~Allanach, C.~G.~Lester, M.~A.~Parker and B.~R.~Webber, \jhep{0009}{2000}{004}.
C.~G.~Lester, M.~A.~Parker and M.~J.~White, \jhep{0601}{2006}{080}.
B.~K.~Gjelsten, D.~J.~Miller and P.~Osland, \jhep{0412}{2004}{003}.
B.~K.~Gjelsten, D.~J.~Miller and P.~Osland, \jhep{0506}{2005}{015}.
\bibitem{glspin}
A.~Alves, O.~Eboli and T.~Plehn, \prd{74}{2006}{095010}.
%
\bibitem{Barger:1992ac}
  V.~D.~Barger, M.~S.~Berger and P.~Ohmann, \prd{47}{1993}{1093}, \hepph{9209232}
    S.~P.~Martin and M.~T.~Vaughn, \prd{50}{1994}{2282}, \hepph{9311340}
%
\bibitem{spira} W. Beenakker, R. Hopker, M. Spira and P. Zerwas,
\npb{492}{1997}{51}.
%
\bibitem{btw} H. Baer, X. Tata and J. Woodside, \prd{42}{1990}{1568}
and \prd{45}{1992}{142}.
%
\bibitem{wbsig}  V.~D.~Barger, A.~L.~Stange and R.~J.~N.~Phillips, 
\prd{45}{1992}{1484}.
%
\bibitem{Wnjet} H.~Baer, V.~D.~Barger and R.~J.~N.~Phillips, \plb{221}{1989}{398}.
%
\bibitem{nuhm} H. Baer, A. Mustafayev, S. Profumo, A. Belyaev 
and X. Tata, \jhep{0507}{2005}{065}. 
%
\bibitem{bcr} H. Baer, C. H. Chen and M. H. Reno, \prd{48}{1993}{5168}.
%
\bibitem{ssdl}
V.~D.~Barger, W.~Y.~Keung and R.~J.~N.~Phillips, \prl{55}{1985}{166}.
R.~M.~Barnett, J.~F.~Gunion and H.~E.~Haber, \plb{315}{1993}{349}.
%
\bibitem{mll} H. Baer, K. Hagiwara and X. Tata, \prd{35}{1987}{1598};
H. Baer, D. Dzialo-Karatas and X. Tata, \prd{42}{1990}{2259};
H. Baer, C. Kao and X. Tata, \prd{48}{1993}{5175};
H. Baer, C. H. Chen, F. Paige and X. Tata, \prd{50}{1994}{4508}. 
%
\bibitem{string} See {\it e.g.} I.~Antoniadis, K.~Benakli, A.~Delgado, M.~Quiros and M.~Tuckmantel,
\npb{744}{2006}{156} and B.~S.~Acharya, K.~Bobkov, G.~L.~Kane, P.~Kumar and J.~Shao,
\hepth{0701034}.
%
\bibitem{splitss} A. Arkani-Hamed and S. Dimopoulos, \jhep{0506}{2005}{073}.
%
\end{thebibliography}
\end{document}